\documentclass[aps,pre,twocolumn,showpacs,preprintnumbers,superscriptaddress,floatfix]{revtex4}
\usepackage{amsmath,amssymb}
\usepackage{graphicx,epsfig}
\usepackage{dcolumn}
\usepackage{bm}
\usepackage{color}

\begin{document}

\title{Boundary-induced abrupt transition in the symmetric exclusion process}

\author{Apoorva Nagar}
\affiliation{School of Physics, Korea Institute for Advanced Study, Seoul 130-722, Korea}

\author{Meesoon Ha}
\affiliation{Department of Physics, Korea Advanced Institute of Science and Technology,
Daejeon, 305-701, Korea}

\author{Hyunggyu Park}
\affiliation{School of Physics, Korea Institute for Advanced Study, Seoul 130-722, Korea}

\date{\today}

\begin{abstract}
We investigate the role of the boundary in the symmetric simple exclusion process with competing nonlocal and local hopping events. With open boundaries, the system undergoes a first order phase transition from a finite density  phase to an empty road  phase as the nonlocal hopping rate increases. Using a cluster stability analysis, we determine the location of such an abrupt nonequilibrium phase transition, which agrees well with numerical results. Our cluster analysis provides a physical insight into the mechanism behind this transition. We also explain why the transition becomes discontinuous in contrast to the case with periodic boundary conditions, in which the continuous phase transition has been observed.
\end{abstract}

\maketitle

\section{INTRODUCTION}
\label{intro}

The study of simple lattice models has emerged as an effective way to understand nonequilibrium steady states and phase transitions~\cite{schmittmann}. In particular, the simple exclusion process~\cite{ligget,derrida} and the zero range process~\cite{spitzer,evans}, owing to their simplicity and richness, have emerged as paradigms for nonequilibrium systems. When such processes are considered with periodic boundary conditions, a phase transition can occur either due to dynamical features such as particle jump rates~\cite{kafri,grosskinsky}, chipping and aggregation~\cite{satya,rajesh}, or due to the presence of disorder on the lattice~\cite{lebowitz,ha2002PGM,barma}. On the other hand, the presence of open boundaries, where particles can be inserted and removed, is significant enough to lead to boundary-induced phase transitions~\cite{krug,ha2003SB}. The phase diagram is well understood for the basic version of the asymmetric simple exclusion process with open boundaries. Besides its analytic solutions~\cite{derrida}, the underlying physical mechanism is also well understood in terms of the group velocity~\cite{ha2002PGM} as well as the domain wall velocity and collective velocity~\cite{kolomeisky}. Recently it was reported that some additional dynamical features such as evaporation and deposition~\cite{parmeggiani} or nonlocal hopping~\cite{ha} lead to new kinds of phases and phase transitions.

In this paper, we study the phase transition in a generalized version of the one-dimensional symmetric simple exclusion process (SSEP) with open boundaries where the additional feature of nonlocal hopping is included. By increasing the nonlocal hopping rate (or equivalently decreasing the input rate of particles at the boundaries), the system undergoes a first order phase transition from a finite density (FD) phase into an empty road (ER) phase with zero density via a clustered state at the transition. This is quite different from the dynamic instability transitions found in the totally asymmetric simple exclusion process (TASEP) with nonlocal hopping studied previously~\cite{ha}. In the TASEP variant, the particle clusters are stable and moving with a drift velocity depending on the nonlocal hopping rate in the most part of the FD phase. In our case, the clusters are stationary on average and stable only at the transition. They expand until they fill up the whole system in the FD phase and shrink away in the ER phase where the whole system is empty except for finite-size clusters clinging to the boundaries. At the transition, the particle clusters are distributed over the system with a power-law gap (intercluster distance) distribution.

The periodic-boundary (PB) version of our model was already investigated~\cite{satya} in terms of a mass transport process. There are similarities and also differences, compared to the open-boundary case. In the PB setup, the number of particles is conserved and plays the role of the external parameter. As the particle density decreases, one finds a continuous condensation (phase separation) transition from a homogeneous density state to a state with an infinite stretch of empty sites (macroscopic mass at a site in the language of the mass transport model). The homogeneous density state is basically the same as the FD state in our open-boundary setup except near the boundaries. The same applies to the critical state. However, the  condensed state includes the background particle clusters with the critical gap distribution over a finite fraction of the system besides the macroscopic gap, while the exponential gap distribution only near the boundaries (vanishing fraction of the system) is found in the ER phase.

Based on the cluster dynamics analysis, we show that the particle density controls the stability of a cluster. Clusters are stable only at a specific density against nonlocal hopping events, which determines the transition point. High-density clusters expand mainly by local hopping events, while low-density clusters shrink easily by nonlocal hopping events. We claim that two seemingly different transitions in both boundary setups are caused by the same mechanism based on the cluster stability. We will also argue that the difference in the nature of the condensed and ER phase is simply due to the presence of possible exits for particles through the open boundaries.

The conventional SSEP has been well studied with both periodic and open boundary conditions~\cite{derrida1}. With periodic or symmetric open boundary setups, the SSEP is an equilibrium system with no phase transitions. Although our dynamical rule in the bulk and at the boundaries does not break the left-right symmetry present in the SSEP, the extra feature of nonlocal hopping drives the system out of equilibrium and leads to a significant change in the steady-state nature for both boundary setups. As a result, contrary to the conventional SSEP with open boundaries, our modified version exhibits an abrupt nonequilibrium transition on varying the input rates at boundaries and the nonlocal hopping rate.

This paper is organized as follows: In Sec.~\ref{phase diagram}, we describe our model and present the numerical results for the phase diagram, clustering, gap distributions, density profile, and finite-size properties. In Sec.~\ref{cluster analysis}, we develop a cluster dynamics analysis at the mean-field level. The cluster stability determines the transition line, which is compared with numerical results. Finally, we discuss the physical origin of this abrupt transition as well as the difference in the periodic version. We conclude the paper in Sec.~\ref{conclusion} with a brief summary.

\begin{figure}[b]
  \centering
  \includegraphics[height=7cm,angle=-90]{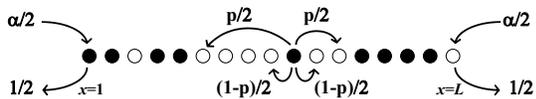}
  \caption{Dynamic processes of the modified SSEP}
  \label{dynamics}
\end{figure}
\begin{figure}[t]
  \centering
  \includegraphics[width=0.975\columnwidth,angle=0]{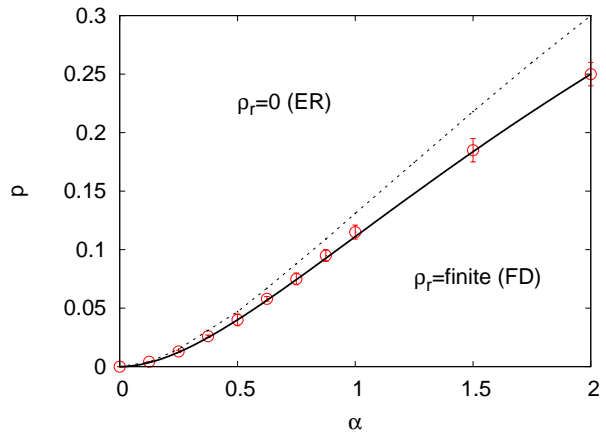}
  \caption{(Color online) $\alpha-p$ phase diagram. The symbols with error bars are obtained from our numerical data and the dashed line is estimated by the cluster mean-field prediction. To guide the eyes, a solid line is drawn through the data points.}
  \label{phase}
\end{figure}
\begin{figure}[b]
  \centering
  \includegraphics[width=0.975\columnwidth,angle=0]{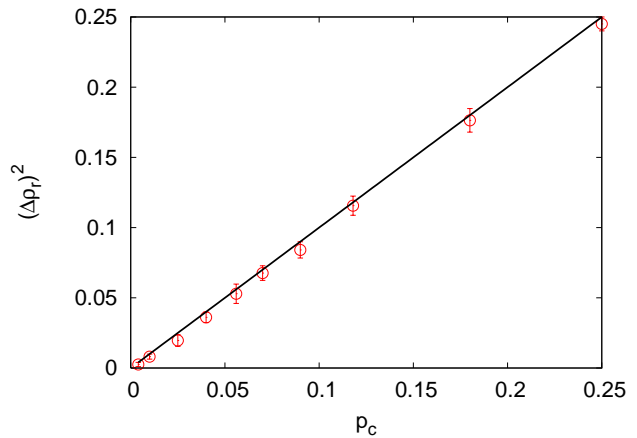}
  \caption{(Color online) Discontinuous jump $\Delta\rho_r$ at the transition $p=p_c$.  The solid line is drawn with $\left(\Delta\rho_r\right)^2=p_c$.}
  \label{jump}
\end{figure}%

\section{The model and Numerical results}
\label{phase diagram}

Consider a chain of length $L$ with sites, $1\le x\le L$, that can either be empty or occupied by a particle with hard core repulsion, thus the occupancy $n_x=0,1$. The chain is open and in contact with two reservoirs at $x=0$ and $x=L+1$, respectively. The evolution rule for our process is as follows: Select a site randomly including two reservoirs ($x=0,\ldots,L+1$). If the site $x$ is occupied by a particle (or one of the reservoirs), we attempt to move the particle. The move is successful only if the target site is unoccupied. The particle either tries to move to the nearest-neighboring site to the left(right) with probability $(1-p)/2$ (local hopping), or it tries to jump to the unoccupied site directly behind the nearest occupied site in the left(right) direction with probability $p/2$ respectively (nonlocal hopping); see Fig.~\ref{dynamics}. In case the chain is completely empty in the chosen direction, the particle jumps all the way into the reservoir. If one of the reservoir sites, $x=0~{\rm or}~L+1$, is selected, a particle tries to jump on the nearest-neighboring site ($x=1~{\rm or}~L$) with probability $\alpha/2$. Nonlocal hopping events from the reservoirs are not allowed in our setup. The particles at the left and right edge of the chain leave the system with probability $1/2$.

\subsection{Phase diagram}

Figure~\ref{phase} shows the phase diagram obtained from our Monte Carlo simulations with road lengths up to $L=4096$ as well as the cluster mean-field (MF) prediction which will be discussed in Sec.~\ref{cluster analysis}. At $p=0$, the process reduces to the conventional SSEP where the stationary state is well known to be the reservoir-controlled uniform phase which we shall call the finite density (FD) phase. Here, the bulk (road) density $\rho_r$ is known exactly as $\rho_r=\alpha/(1+\alpha)$. The FD phase extends into $p>0$ for all $\alpha$ with $\rho_r$ decreasing slowly with $p$. For strong nonlocal hopping (large $p$), the stationary state changes into the empty road (ER) phase with $\rho_r=0$ with finite-size {\em mother} clusters clinging to reservoirs.

The phase transition from the FD phase into the ER phase turns out to be always discontinuous with a jump of $\Delta\rho_r\simeq \sqrt{p_c}$ at the transition $p=p_c$ (Fig.~\ref{jump}). This implies that this instability transition is caused by competition of the bulk density against the nonlocal hopping rate. The boundary parameters such as the input rate $\alpha$ only play an implicit role in this transition via determining the bulk density . This suggests that our open-boundary version should be very similar, if not identical, to the periodic-boundary version which is controlled directly by the bulk density and the nonlocal hopping rate. In fact, the phase boundary is known exactly for the periodic-boundary version as $p_c=\rho_r^2$~\cite{satya}, which is also likely to determine the phase boundary in our open-boundary setup. Our numerical results strongly support this. However, it is not trivial to find the bulk density $\rho_r$ as a function of $\alpha$ and $p$. We estimate this at the mean-field level in Sec.~\ref{cluster analysis} and the corresponding phase boundary is shown in Fig.~\ref{phase}, which agrees reasonably well with numerical results.

\begin{figure}
  \centering
  \includegraphics[width=0.975\columnwidth,angle=0]{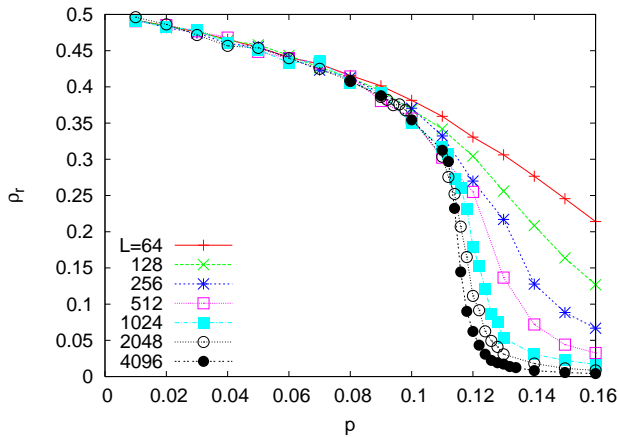}
  \caption{(Color online) The bulk density $\rho_r$ versus $p$ at $\alpha=1$ for various system sizes up to $L=4096$.}
  \label{density}
\end{figure}

\begin{figure}
  \centering
  \includegraphics[width=0.975\columnwidth,angle=0]{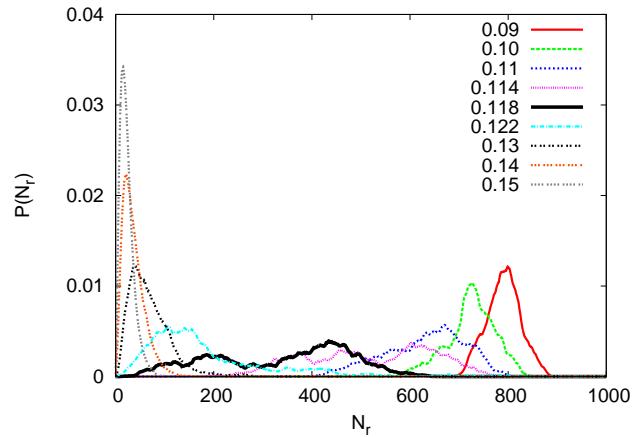}
  \caption{(Color online) Probability distribution $P(N_r)$ for finding $N_r$ particles on the road at $\alpha=1$ for various $p=0.09,~0.10,~...,~0.15$ (from right to left) at $L=2048$.
  The FD-ER transition occurs at $p_c\simeq 0.118(4)$.}
  \label{probden}
\end{figure}
\begin{figure}
  \centering
  \includegraphics[width=0.975\columnwidth,angle=0]{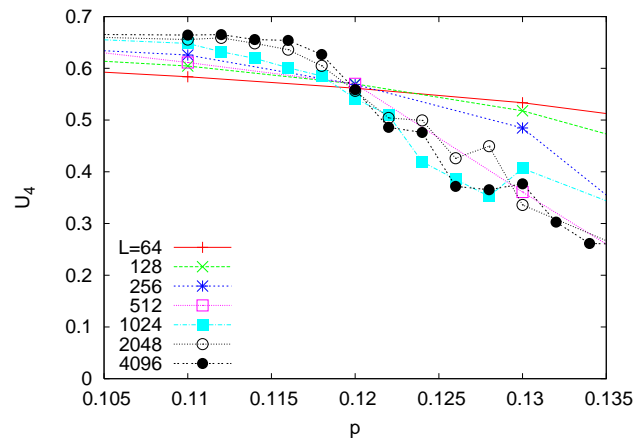}
  \caption{(Color online) The Binder cumulants versus $p$ at $\alpha=1$ for various system sizes up to $L=4096$. The crossing points converge to the point at $p_c\simeq 0.118(2)$ and $U_4(p_c)\simeq 0.56(3)$.}
  \label{binder}
\end{figure}

To establish the phase diagram accurately, we measure various quantities in the steady state such as the particle density $\rho_r=\langle n_r\rangle$ with $n_r\equiv \frac{1}{L}\sum_{x=1}^L n_x$, the density fluctuations $\chi_2=L(\langle n_r^2\rangle-\langle n_r\rangle^2)$, and the Binder cumulant $U_4=1-\langle n_r^4\rangle/(3\langle n_r^2\rangle^2)$. In Fig.~\ref{density}, the bulk density $\rho_r$ is plotted against $p$ at $\alpha=1$. The systematic shift in the data with increasing size clearly indicates that, in the infinite-size limit, there will be a discontinuous jump in $\rho_r$ at $p=p_c$. Figure~\ref{probden} shows the evolution of the probability distribution $P(N_r)$ for finding $N_r=\rho_r L$ particles on the road across the transition. At the transition estimated as $p_c\simeq 0.118(4)$, we see an abrupt change into a broad distribution, which is again a signature of a discontinuous transition.

The instability threshold $p_c$ can be accurately estimated by the crossing points of the Binder cumulant $U_4$. Fig.~\ref{binder} shows a nice convergence of the crossing points to the transition point $p_c\simeq 0.118(2)$. The Binder cumulant $U_4$ takes a value of $2/3$ in the FD phase and $0$ in the ER phase. At the transition, $U_4(p_c)\simeq 0.56(3)$. Note that the distributions in the ER phase and at the transition are slightly different from those for the TASEP variant studied in ~\cite{ha}. We also observe that the density fluctuation $\chi_2$ becomes maximum at the transition as expected (not shown here). All other data for various values of $\alpha$ show a similar behavior to the $\alpha=1$ case. Numerical data for the phase boundary and the density jump are shown in Figs.~\ref{phase} and~\ref{jump}.

\begin{figure*}[t]
  \centering
  \includegraphics[width=16cm,angle=0]{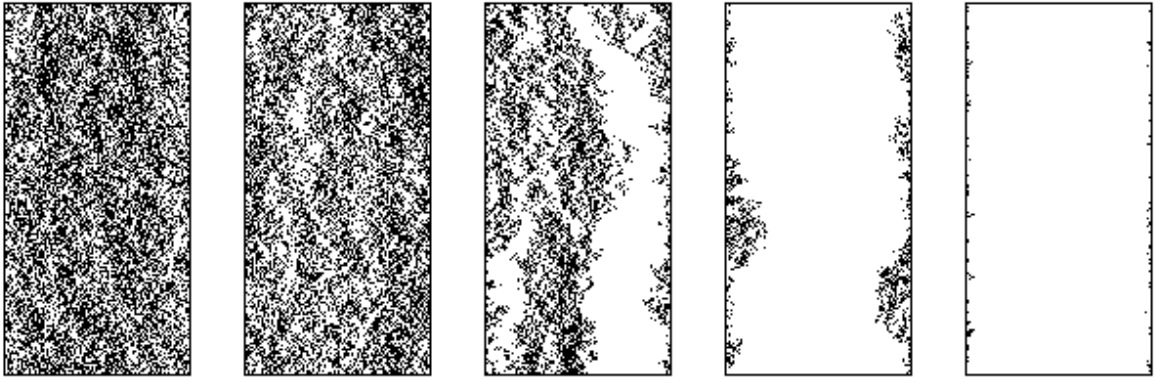} 
  \caption{Steady-state space(horizontal)-time(vertical) density profiles for $L=256$ at $\alpha=1$. From left to right, $p=0,~ 0.05,~ 0.12~(\approx p_c),~ 0.2,~0.5$. The steady state is homogeneous with the finite bulk density for $p<p_c$ (FD phase) and there is a cluster formation at $p_c$. The last two plots show typical patterns in the ER phase ($p>p_c$). }
  \label{timeplot}
\end{figure*}
\begin{figure}
  \centering
  \includegraphics[width=0.975\columnwidth,angle=0]{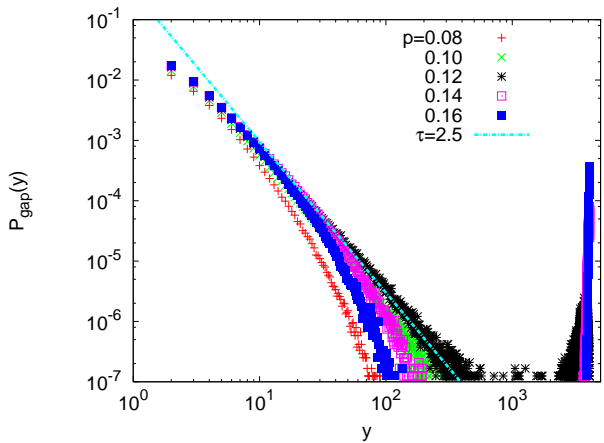}
  \caption{(Color online) Probability distribution $P_{\rm gap}(y)$ for the gap size $y$ on the road for $L=4096$
  at $\alpha=1$ ($p_c\approx 0.12$).}
  \label{gapdist}
\end{figure}
\begin{figure}
  \centering
  \includegraphics[width=0.975\columnwidth,angle=0]{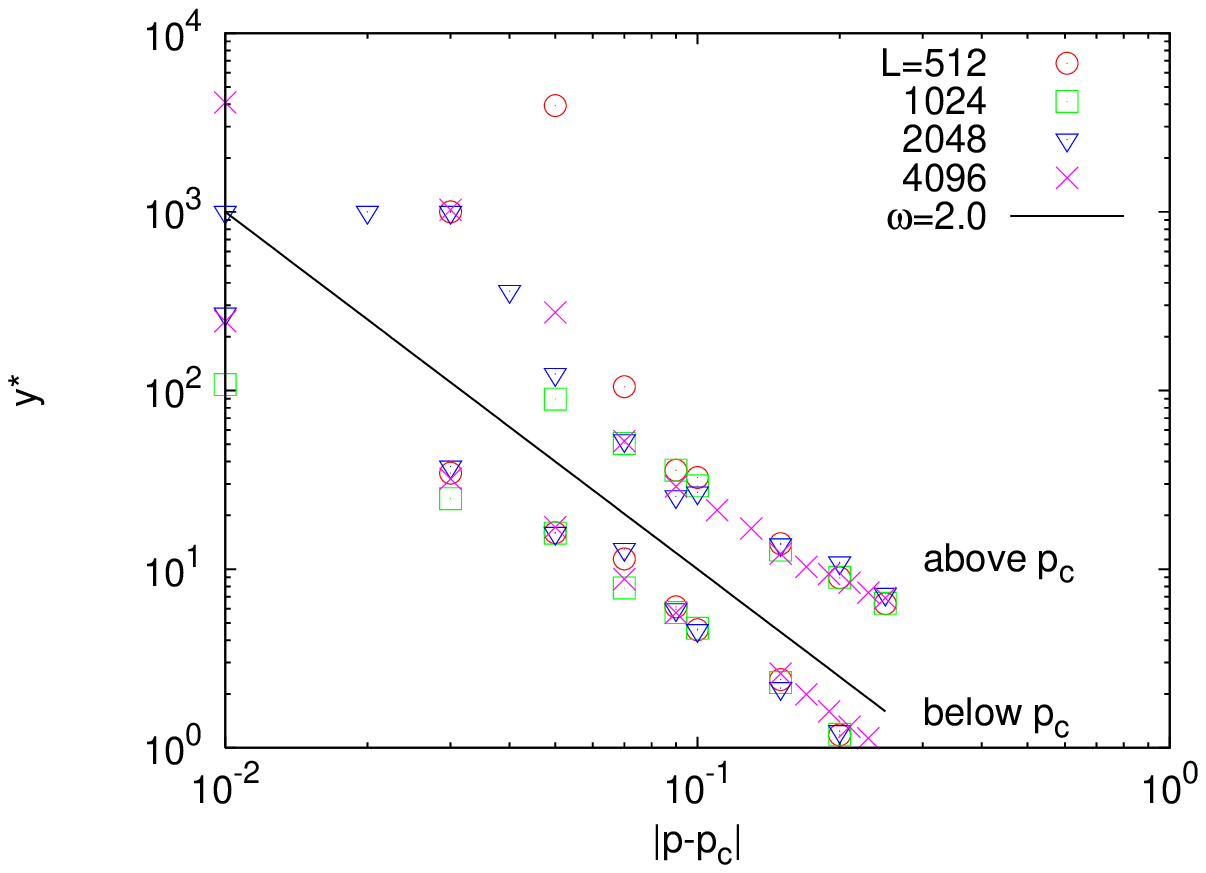}\\
  \includegraphics[width=0.975\columnwidth,angle=0]{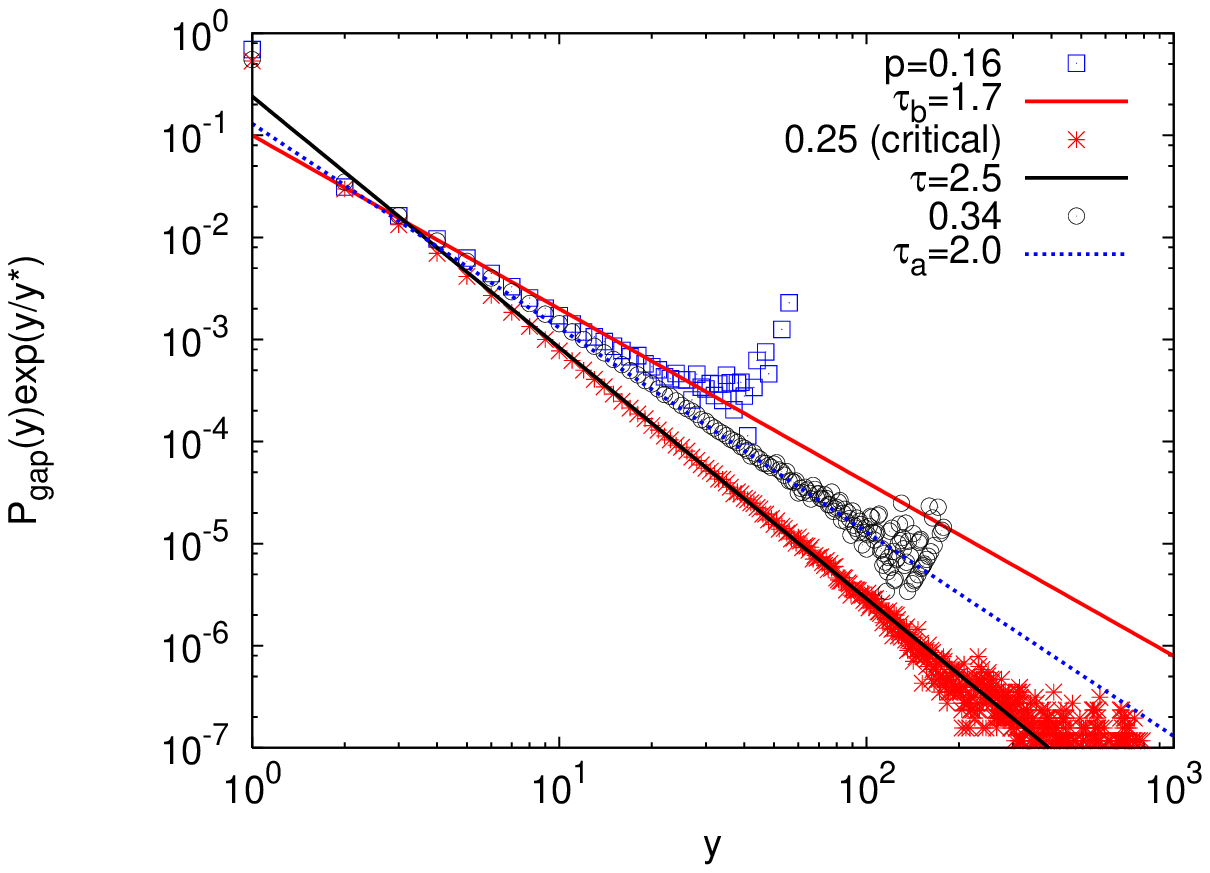}
  \caption{(Color online) Scaling behavior of the cutoff gap size $y^*$ near $p_c\approx 0.25$ (upper) and the power-law tail test for $P_{\rm gap}(y)$  for $L=4096$ (lower) at $\alpha=2$.}
  \label{gap-analysis}
\end{figure}
\begin{figure}[t]
 \centering
  \includegraphics[width=0.975\columnwidth,angle=0]{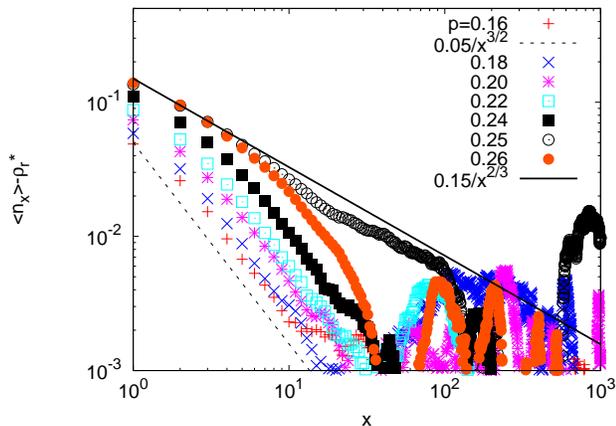}
  \caption{(Color online) Density profiles at various values of $p$ $(p_c\approx 0.25)$ with $\alpha=2$ and $L=4096$.}
  \label{profile}
\end{figure}
\begin{figure}[b]
  \centering
  \includegraphics[width=0.975\columnwidth,angle=0]{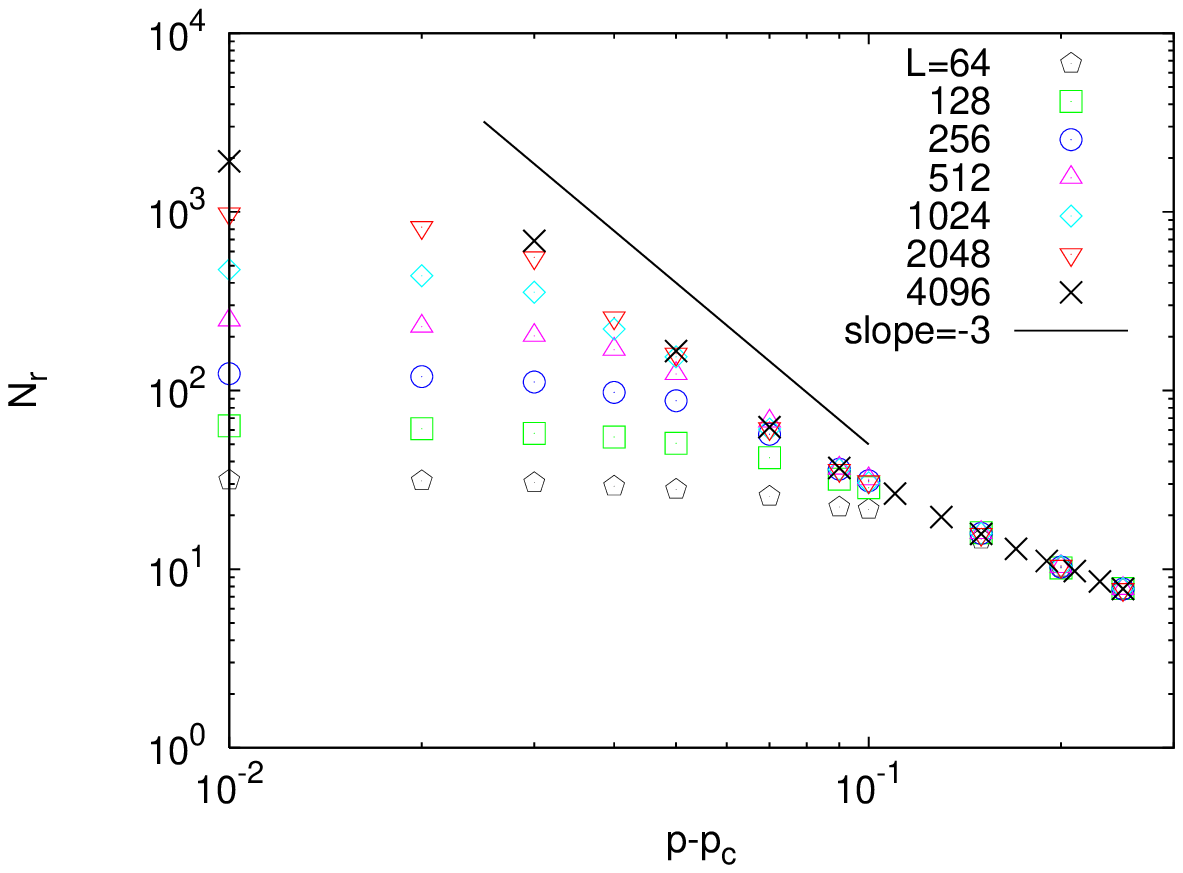}
  \caption{(Color online) Double logarithmic plots of $N_r=\rho_rL$ against $p-p_c$ at $\alpha=2$
   for various system sizes up to $L=4096$. Its asymptotic behavior near $p_c$ from the ER phase side seems to be
   $N_r\sim (p-p_c)^{-\zeta}$ with $\zeta\approx 3$.}
  \label{Nr-zeta}
\end{figure}

\subsection{Gap size distribution}

The spatiotemporal patterns in the steady state provide some insights into the phases and the phase transition, see Fig.~\ref{timeplot}. The FD phase shows a homogeneous distribution of particles, while in the ER phase one big gap appears with small clusters near the boundaries. At the transition, particle clusters are formed with inter-cluster gaps of various sizes.

We investigate the distribution of the gap size (the size of empty stretches) numerically, which is shown in Fig.~\ref{gapdist}. From these data, we conclude that the probability distribution $P_{\rm gap}(y)$ for gap size $y$
behaves as
\begin{equation}
P_{\rm gap}(y)\sim \left\{
\begin{array}{ll}
y^{-\tau_b} e^{-y/y^*_{b}} & \mbox{for}\ p<p_c ,\\
y^{-\tau} & \mbox{at}\ \ p=p_c ,\\
y^{-\tau_a} e^{-y/y^*_{a}}+\frac{1}{L\rho_r} \delta(y-aL) & \mbox{for}\ p>p_c .
\end{array}
\right.
\label{eq-gap}
\end{equation}

In the FD phase ($p<p_c$),  $P_{\rm gap}(y)$ decays exponentially as expected  for a  homogeneous density state. In the ER phase ($p>p_c$),  it decays exponentially for small $y$ as in the FD phase, but in addition a $\delta$-function peak appears at $y=aL$ with $a=1-\rho_r$ approaching $1$ in the $L\rightarrow\infty$ limit. The $\delta$-function peak represents the macroscopic gap in the center of the system seen in Fig.~\ref{timeplot} and the exponential part describes the structure of finite mother clusters clinging to the reservoirs. Typical gap sizes in the FD phase and also in the mother clusters in the ER phase diverge when approaching the transition from below and above; $y^*_{a,b}\sim |p_c-p|^{\omega_{a,b}}$ with $\omega_a\approx \omega_b\approx 2.0$. The multiplicative power-law exponents are also numerically estimated as $\tau_a\simeq 2.0(2)$ and $\tau_b\simeq 1.7(3)$. At the transition ($p=p_c$), $P_{\rm gap}(y)$ decays as a power law with the exponent $\tau \simeq 2.5(1)$ (see Figs.~\ref{gapdist} and~\ref{gap-analysis}).

Now we compare our results with those in the periodic-boundary (PB) version~\cite{satya}. The gap distributions in the FD phase and at the transition are found to be almost identical to those in the PB version. The numerical values of the exponents $(\omega_b, \tau_b, \tau)$ are very close to the mean-field predictions of $(2, 3/2, 5/2)$~\cite{satya}. The main difference lies in the ER phase where an exponential decay is observed with a $\delta$-function while, in the PB setup, the $\delta$-function is accompanied by a power law with the same exponent $\tau$ as at criticality in the condensed phase. The exponential decay involves two new exponents $\omega_a\approx 2.0$ and $\tau_a=2.0(2)$ in the ER phase. Note that $\tau_a$ seems different from $\tau_b$ and $\tau$. Moreover, the $\delta$-function in our case has a fixed location while the location of the $\delta$-function varies as a function of the control parameter $\rho$ (the particle density) in the PB setup. We shall reflect on this difference in the next section.

As shown in our spatiotemporal plots, Fig.~\ref{timeplot},
there is a formation of clusters at $p=p_c$ and the power-law gap distribution is an indicator of such clusters.
In the ER phase, all clusters dissolve away except two mother clusters near the boundaries. We believe that the reason for this phase transition lies in the instability of clusters. Moreover, the clusters could be highly sensitive to the open boundaries with which the system would be unable to sustain the clusters inside. We will discuss this point via a cluster stability analysis in the next section.

\subsection{Density profile and finite-size property}

We finally discuss the asymptotic behavior of the density profile $\langle n_x\rangle$ as well as the finite-size property of the road density $\rho_r$. Based on numerical data,
we propose
\begin{equation}
\langle n_x \rangle \simeq \left\{
\begin{array}{ll}
\rho_r^{*}(\alpha,p)+Bx^{-\kappa_b} & \mbox{for}\ p<p_c ,\\
\Delta \rho_r^{*}(p_c)+Cx^{-\kappa} & \mbox{at}\ \ p=p_c^- ,\\
A e^{-x/x^*} & \mbox{for}\ p>p_c ,
\end{array}
\right.
\label{eq-rho-x}
\end{equation}
where $\rho_r^{*}$ is the bulk density in the infinite-size limit.
Figure~\ref{profile} shows that the density profile has a power-law tail near the boundary with $\kappa_b\approx 3/2$ in the FD phase; $\kappa\approx 2/3$ at the transition approaching from the FD side; and the exponential tail in the ER phase. This is quite different from that in the ordinary open SSEP where $\rho(x)$ does not have any tail structure and it is uniform when there is the left-right symmetry~\cite{derrida1}.

One can calculate the finite-size property of the road density by integrating the density profile up to size $L$:
\begin{equation}
\rho_r\simeq \left\{
\begin{array}{ll}
\rho_r^{*}(\alpha,p)+\tilde{B}L^{-1}+O(L^{-\kappa_b}) & \mbox{for}\ p<p_c ,\\
\Delta \rho_r^{*}(p_c)+\tilde{C}L^{-\kappa}+O(L^{-1}) & \mbox{at}\ \ p=p_c^- ,\\
\tilde{A}L^{-1} & \mbox{for}\ p>p_c .
\end{array}
\right.
\label{eq-rho}
\end{equation}
In the ER phase, the total number of particles, $N_r(=\rho_rL)$, becomes a $p$-dependent constant $\tilde{A}$ in the infinite-size limit. Approaching the transition point from the ER side, we expect this constant to diverge. Figure~\ref{Nr-zeta} suggests the power-law type divergence as $N_r\sim (p-p_c)^{-\zeta}$ with $\zeta\approx 3.0$. One may relate the exponent $\zeta$ to the finite-size-scaling exponent $\phi$ which is termed the crossover exponent in the PB version. The cutoff gap size $y_a^*$ should scale as $\sim N_r^\phi$ close to the transition [see Eq.~(20) in \cite{satya}] and also as $\sim (p-p_c)^{-\omega_a}$ with $\omega_a\approx 2.0$. Therefore we expect $\zeta=\omega_a/\phi$, leading to $\phi\approx 2/3$, which is again consistent with the mean-field result for the PB version~\cite{satya}.

\section{CLUSTER MEAN-FIELD ANALYSIS}
\label{cluster analysis}

\subsection{Cluster stability analysis}

In the TASEP variant studied previously~\cite{ha}, the cluster dynamics analysis provides a key insight to understanding the instability transition and also leads to quite accurate predictions. Here we take a similar approach and focus on the stability of clusters. As can be seen in Fig.~\ref{timeplot} at the transition, there are many floating clusters (free clusters) in the bulk besides two clusters attached to the reservoirs (mother clusters). Assume that the steady state at the transition consists of these clusters.

First, consider the free clusters with the average bulk density $\rho_c$ and the side density $\rho_{_{\rm S}}$ (the side densities at the right and left side of the clusters are expected to be identical for free clusters). The right front of such a cluster moves with a velocity given by
\begin{eqnarray}
u_{_{\rm R}}= \frac{(1-p)}{2}-\frac{p}{2\rho_{_{\rm S}}}-\frac{(1-\rho_{_{\rm S}})}{2}+\frac{p}{2}.
\end{eqnarray}
The first (second) term on the right-hand side (RHS) arises from a local (nonlocal) move of the particle at the front to the right. The cluster front creeps forward by its local move to the right, but shrinks back due to the detachment of the front particle by a nonlocal hopping to the next cluster in front. The third term arises from its move to the left, which is possible only when the target site is vacant (nonlocal hops are ignored inside the clusters). The last term is due to the nonlocal hop of a particle from the next cluster in front, which makes the cluster front advances by one unit.
One can write a similar equation for the left front velocity of the cluster. Since there is the left-right symmetry in the dynamics, our clusters do not drift in either direction ($u_{_{\rm R}}+u_{_{\rm L}}=0$) and the clusters are stable only when $u_{_{\rm R}}=u_{_{\rm L}}=0$. This leads to the conditions:
\begin{eqnarray}
{\rho_{_{\rm S}}^c}=\sqrt{p},
\end{eqnarray}
where ${\rho_{_{\rm S}}^c}$ is the critical side density for the stable clusters.

Note that, for given $p$, the clusters are stable only when their side density is precisely equal to the critical density  and they are extremely sensitive to a slight change of their side density. If the side density increases temporarily by fluctuations, the front velocity becomes positive and the clusters start to expand and merge into each other until they fill up the road, if the reservoirs supply particles sufficiently into the free clusters through the mother clusters. In the opposite case, the clusters start to shrink and vanish, leading to the appearance of a big empty stretch on the road, if the mother clusters do not completely block the escape of particles to the reservoirs.

Now, consider the mother clusters. The right front of the mother cluster attached to the left reservoir is free to expand or contract. Similar to the free cluster analysis, we find the front velocity as
\begin{eqnarray}
u_{_{\rm F}}= \frac{(1-p)}{2}-\frac{p}{2\rho_{_{\rm F}}}-\frac{(1-\rho_{_{\rm F}})}{2}+\frac{p}{2},
\end{eqnarray}
where $\rho_{_{\rm F}}$ is the front density of the mother cluster. For the mother cluster to be stable, this velocity should be zero, which leads to the condition:
\begin{eqnarray}
\rho_{_{\rm F}}^c=\sqrt{p}.
\end{eqnarray}
Similar to the free clusters, the stability of  the mother clusters is extremely fragile to density fluctuations and their critical front density is exactly the same as that of the free clusters.

Since the mother cluster is in a sense a buffer between the boundary and the bulk, it has a continuously varying density with a higher value at the boundary and a lower value into the bulk. If this bulk density is higher than the critical density $\rho_{_{\rm F}}^c$, the mother cluster expands and merges with bulk free clusters until they fill up the whole road uniformly. In the opposite situation, it shrinks and retreats close to the boundary until the front density becomes equal to the critical density. This short mother cluster should be very unstable and fluctuate wildly. In this case, the bulk free clusters may escape away from the system via nonlocal hopping whenever the mother clusters vanish intermittently. Therefore we expect, in the steady state, only two short mother clusters with a big macroscopic empty road in the bulk. When the bulk density is precisely equal to the critical density, the mother cluster is stable. Also, due to the left-right symmetry in dynamics, the side density of the free clusters is expected to be the same as the front density of the mother cluster: $(\rho_{_{\rm F}}=\rho_{_{\rm S}})$. Therefore the free clusters are also stable in this case and we expect a cluster fluid. Next, we calculate the density profile of the mother cluster to determine the bulk density as function of the external parameters $p$ and $\alpha$ .

\subsection{Density profile for mother clusters }

The stability conditions of clusters provide a criterion for the phase transition in terms of the front or side density of clusters and the nonlocal hopping parameter $p$. We would now like to bring the input rate $\alpha$ into the picture, which will determine the particle density profile of the mother cluster. In the FD phase (small $p$), the steady state has no clustering with finite bulk density. In this regime, we may use a simple mean-field (MF) analysis to determine the density profile. Approaching the transition, the system becomes correlated and the MF results may not be trustworthy. However, the correlation is finite at the discontinuous transition, so the deviation from the MF results is expected to be  small.

The density of the mother cluster decreases from a higher value near the reservoir and settles to a constant value in the bulk. This bulk value $\rho_b$ is what we are interested in: At the transition, this value is equal to the critical density; $\rho_b (p,\alpha)=\rho^c=\sqrt{p}$.

To determine the MF density profile, let us consider the current at the bulk bond between sites $x$ and $x+1$ ($x=1,\ldots,L-1$). The average current flowing to the right is given by
\begin{eqnarray}
J_{x+1/2}^{\rm R}=\frac{\langle n_x v_{x+1} \rangle}{2}+\frac{p\langle v_x v_{x+1}\rangle}{2}-\frac{pP_0(x+1)}{2},
\end{eqnarray}
where $n_x$ is the occupancy of site $x$ and the vacancy  $v_x=1-n_x$. The first term on the RHS accounts for the local jump from $x$ to $x+1$ and the second term for the nonlocal hop of a particle on a site to the left of $x$. The third term accounts for the case in which there is no such particle at all to the left of $x$, which would not allow any nonlocal hop: $P_{0}(x)$ is given by
\begin{eqnarray}
P_{0}(x)=\langle\prod_{y=1}^{x}v_y \rangle.
\end{eqnarray}
One can similarly write the expression for the left current
\begin{eqnarray}
J_{x+1/2}^{\rm L}=\frac{\langle n_{x+1} v_{x} \rangle}{2}+\frac{p\langle v_x v_{x+1}\rangle}{2}-\frac{pQ_0(x)}{2}
\end{eqnarray}
with $Q_{0}(x)=\langle\prod_{y=x}^{L}v_y \rangle$.

The symmetry in the problem dictates that, in the steady state, the average current through a site should be zero as
\begin{eqnarray}
J_{x+1/2}^R-J_{x+1/2}^L=0.
\label{LRcurrent}
\end{eqnarray}

For $x\ll L/2$, we may take $Q_{0}(x)=0$ in the FD phase. Combining this with Eq.~(\ref{LRcurrent}), we arrive at
\begin{eqnarray}
\langle v_{x+1}  \rangle = \langle v_{x}  \rangle +pP_0(x+1)  .
\end{eqnarray}
Employing the MF approximation such as $P_0(x+1)=P_0(x)\langle v_{x+1}  \rangle$, one can write the recurrence relations for the vacancy profile in the bulk
\begin{eqnarray}
\langle v_{x+1} \rangle = \frac{\langle v_x \rangle}{1-pP_0(x)}.
\label{mfone}
\end{eqnarray}

Now consider the current at the boundary between the left reservoir ($x=0$) and the site $x=1$. One may easily find
\begin{eqnarray}
J_{1/2}^{\rm R}=\frac{\alpha \langle v_1 \rangle}{2},~~~~J_{1/2}^L=\frac{\langle n_1 \rangle}{2}+\frac{p\langle v_1 \rangle}{2}-\frac{pQ_0(1)}{2}.
\end{eqnarray}
Again, the total average current is zero and $Q_0(1)=0$ for the FD phase and also at the transition. Thus, we get
\begin{eqnarray}
\langle v_1 \rangle =\frac{1}{1+\alpha-p}.
\label{mftwo}
\end{eqnarray}
Using Eqs.~(\ref{mfone}) and~(\ref{mftwo}), we determine the density profile iteratively. The density decreases from a high value near the reservoir and saturates to a bulk value $\rho_b(p,\alpha)$. Using the stability condition $\rho_b=\sqrt{p}$, the phase boundary $p_c=p_c(\alpha)$ is evaluated, which is shown in Fig.~\ref{phase}.  For small $p$, the MF prediction agrees very well with the Monte Carlo data.  As $p$ increases, the nonlocal events that induce a strong density-density correlations become important. Thus, the results of our MF analysis show some deviation from the Monte Carlo results.

\subsection{Discussions}

Our cluster stability analysis may be directly applied to the periodic-boundary (PB) version. Of course, the stability condition derived here is not exact due to ignoring nonlocal intracluster hops. Moreover the side or front density can not be equivalent to the given bulk density $\rho_r$ in the PB version. Nevertheless, it is interesting to note that the exact transition line equation $p_c=\rho_r^2$~\cite{satya} coincides with our stability equation.

The picture developed in the cluster stability analysis explains the sudden jump in the density leading to a discontinuous transition, in terms of the formation and dissolution of clusters. Let us compare our case to the PB version, where the phase transition is continuous. The system studied in~\cite{satya} is based on the mass transport process, so it has no restriction on the occupancy of sites and the dynamical rules consist of particle clusters diffusing and merging on contact. In addition, single particles can chip off from clusters and move to the neighboring sites. The boundary conditions are taken to be periodic. This model can be mapped exactly onto the SSEP with nonlocal hopping in the PB setup.

In the SSEP language, it shows a continuous phase transition from the uniform FD phase into the condensed phase with a macroscopic gap. In the condensed phase (corresponding to the ER phase in our model), the gap distribution is still critical (the same as at the transition) with a trivial macroscopic gap. This may also be understood with our picture of the free cluster stability. Note that there are only free clusters at the transition in the PB setup. Going into the condensed phase by lowering the total density, the clusters are diluted and thus become unstable by emitting particles away. However, in contrast to our model with open boundaries, particles can not escape the system in the PB version. Instability and fluctuations may drive particles into a finite (but macroscopic) segment of the system where the local density is equal to the critical bulk density. Then this finite segment can maintain its stability by shooting particles via nonlocal hopping through a macroscopic empty road. As the particle density in this segment is critical, the gap distribution should be exactly the same as that at the transition with only one extra macroscopic gap. In our open boundary setup, the mother clusters become highly unstable and the free clusters escape away through the boundaries. Thus, the resulting gap distribution become exponential.

The periodic and open boundary conditions discussed here define different ensembles such as canonical and grand-canonical ensembles where the number of particles is conserved or controlled by such external parameters as the input rate. In the equilibrium systems, it is well known that two ensembles are equivalent in the thermodynamic limit. In nonequilibrium systems, this ensemble equivalence does not need to hold in general. Our study provides a simple example in which the open boundary is crucial to the nature of the instability transition.

\section{CONCLUSIONS}
\label{conclusion}

In this paper, we studied how the nonequilibrium system is affected by different boundary conditions. In the case of the symmetric simple exclusion process (SSEP) with nonlocal hopping in one spatial dimension, we showed that the presence of open boundaries induces a boundary-induced abrupt transition. We developed the cluster stability analysis to explain this abrupt transition successfully, which was compared and contrasted to the system without boundaries (closed chain). We found that the cluster stability governs the underlying physical mechanism of instability-induced phase transitions, and we discussed the ensemble equivalence in the generalized SSEP. It is also interesting to test the ensemble equivalence in the generalized asymmetric exclusion process with the same type of nonlocal hopping events studied in~\cite{ha}, which will be discussed elsewhere.

\section*{Acknowledgements}
This work was supported by the BK21 project and Acceleration Research (CNRC) of MOST/KOSEF. Computation was carried out using KIAS supercomputers. H.P. would like to acknowledge the kind hospitality of the Kavli Institute for Theoretical Physics China (KITPC), where the most part of this manuscript was completed.


\begin{thebibliography}{40}
\expandafter\ifx\csname natexlab\endcsname\relax\def\natexlab#1{#1}\fi
\expandafter\ifx\csname bibnamefont\endcsname\relax
  \def\bibnamefont#1{#1}\fi
\expandafter\ifx\csname bibfnamefont\endcsname\relax
  \def\bibfnamefont#1{#1}\fi
\expandafter\ifx\csname citenamefont\endcsname\relax
  \def\citenamefont#1{#1}\fi
\expandafter\ifx\csname url\endcsname\relax
  \def\url#1{\texttt{#1}}\fi
\expandafter\ifx\csname urlprefix\endcsname\relax\def\urlprefix{URL }\fi
\providecommand{\bibinfo}[2]{#2}
\providecommand{\eprint}[2][]{\url{#2}}

\bibitem[{\citenamefont{Schmittmann}(1995)}]{schmittmann}
\bibinfo{author}{\bibfnamefont{B.} \bibnamefont{Schmittmann}} \bibnamefont{and~}\bibinfo{author}{\bibfnamefont{R.~K.~P.} \bibnamefont{Zia}}, \emph{\bibinfo{title}{Statistical Mechanics of Driven Diffusive Systems}}, \emph{\bibinfo{title}{Academic, London}}, (\bibinfo{year}{1995});

\bibitem[{\citenamefont{Ligget}(1983)}]{ligget}
\bibinfo{author}{\bibfnamefont{T.~M.} \bibnamefont{Ligget}}, \emph{\bibinfo{title}{Interacting Particle Systems}}, \emph{\bibinfo{title}{Springer-Verlag, New York}}, (\bibinfo{year}{1983}); \bibinfo{author}{\bibfnamefont{T.~M.} \bibnamefont{Ligget}}, \emph{\bibinfo{title}{Stochastic Interacting Systems: Contact, Voter and Exclusion Process}}, \emph{\bibinfo{title}{Springer-Verlag}}, (\bibinfo{year}{1999}).

\bibitem[{\citenamefont{Derrida}(1993)}]{derrida}
\bibinfo{author}{\bibfnamefont{B.} \bibnamefont{Derrida}}, \bibinfo{author}{\bibfnamefont{M.~R.} \bibnamefont{Evans}},\bibinfo{author}{\bibfnamefont{V.} \bibnamefont{Hakim}} \bibnamefont{and} \bibinfo{author}{\bibfnamefont{V.} \bibnamefont{Pasquier}},
\bibinfo{journal}{J. Phys. A} \textbf{\bibinfo{volume}{26}}, \bibinfo{pages}{1493} (\bibinfo{year}{1993}); \bibinfo{author}{\bibfnamefont{G.} \bibnamefont{Sch{\"u}tz}} \bibnamefont{and} \bibinfo{author}{\bibfnamefont{E.} \bibnamefont{Domany}}, \bibinfo{journal}{J. Stat. Phys.} \textbf{\bibinfo{volume}{72}},
\bibinfo{pages}{277} (\bibinfo{year}{1993}); \bibinfo{author}{\bibfnamefont{C.} \bibnamefont{MacDonald}}, \bibinfo{author}{\bibfnamefont{J.} \bibnamefont{Gibbs}} \bibnamefont{and} \bibinfo{author}{\bibfnamefont{A.} \bibnamefont{Pipkin}}, \bibinfo{journal}{Biopolymers} \textbf{\bibinfo{volume}{6}}, \bibinfo{pages}{1} (\bibinfo{year}{1968})

\bibitem[{\citenamefont{Spitzer}(1970)}]{spitzer}
\bibinfo{author}{\bibfnamefont{F.} \bibnamefont{Spitzer}},
\bibinfo{journal}{Adv. Math.} \textbf{\bibinfo{volume}{5}}, \bibinfo{pages}{246} (\bibinfo{year}{1970}).

\bibitem[{\citenamefont{Evans}(2000)}]{evans}
\bibinfo{author}{\bibfnamefont{M. R.} \bibnamefont{Evans}},
\bibinfo{journal}{Braz. J. Phys.} \textbf{\bibinfo{volume}{30}}, \bibinfo{pages}{42} (\bibinfo{year}{2000}); \bibinfo{author}{\bibfnamefont{M.~R.} \bibnamefont{Evans}}, \bibnamefont{and} \bibinfo{author}{\bibfnamefont{T.} \bibnamefont{Hanney}}, \bibinfo{journal}{J. Phys. A: Math. Gen.} \textbf{\bibinfo{volume}{38}}, \bibinfo{pages}{R195}
(\bibinfo{year}{2005}).

\bibitem[{\citenamefont{Kafri}(2002)}]{kafri}
\bibinfo{author}{\bibfnamefont{Y.} \bibnamefont{Kafri}},
\bibinfo{author}{\bibfnamefont{E.} \bibnamefont{Levine}},
\bibinfo{author}{\bibfnamefont{D.} \bibnamefont{Mukamel}},
\bibinfo{author}{\bibfnamefont{G.~M.} \bibnamefont{Sch{\"u}tz}},
\bibnamefont{and} \bibinfo{author}{\bibfnamefont{J.} \bibnamefont{T{\"o}r{\"o}k}},
\bibinfo{journal}{Phys. Rev. Lett.} \textbf{\bibinfo{volume}{89}}, \bibinfo{pages}{035702}
(\bibinfo{year}{2002}).

\bibitem[{\citenamefont{Grosskinsky}(2003)}]{grosskinsky}
\bibinfo{author}{\bibfnamefont{S.} \bibnamefont{Grosskinsky}}, \bibnamefont{and~} \bibinfo{author}{\bibfnamefont{G.~M.} \bibnamefont{Sch{\"u}tz}}, \bibinfo{journal}{J. Stat. Phys.} \textbf{\bibinfo{volume}{113}}, \bibinfo{pages}{389}
(\bibinfo{year}{2003}).

\bibitem[{\citenamefont{Satya}(2000)}]{satya}
\bibinfo{author}{\bibfnamefont{S.~N} \bibnamefont{Majumdar}}, \bibinfo{author}{\bibfnamefont{S.} \bibnamefont{Krishnamurthy}}, \bibnamefont{and} \bibinfo{author}{\bibfnamefont{M.} \bibnamefont{Barma}},
\bibinfo{journal}{Phys. Rev. Lett.} \textbf{\bibinfo{volume}{81}}, \bibinfo{pages}{3691} (\bibinfo{year}{1998});\bibinfo{author}{\bibfnamefont{S.~N} \bibnamefont{Majumdar}}, \bibinfo{author}{\bibfnamefont{S.} \bibnamefont{Krishnamurthy}}, \bibnamefont{and} \bibinfo{author}{\bibfnamefont{M.} \bibnamefont{Barma}},
\bibinfo{journal}{J. Stat. Phys.} \textbf{\bibinfo{volume}{99}}, \bibinfo{pages}{1} (\bibinfo{year}{2000}); \bibinfo{author}{\bibfnamefont{R.} \bibnamefont{Rajesh}} \bibnamefont{and} \bibinfo{author}{\bibfnamefont{S.~N} \bibnamefont{Majumdar}}, \bibinfo{journal}{Phys. Rev. E} \textbf{\bibinfo{volume}{63}}, \bibinfo{pages}{036114} (\bibinfo{year}{2001}).

\bibitem[{\citenamefont{Rajesh}(2002)}]{rajesh}
\bibinfo{author}{\bibfnamefont{R.} \bibnamefont{Rajesh}}
\bibnamefont{and} \bibinfo{author}{\bibfnamefont{S.} \bibnamefont{Krishnamurthy}},
\bibinfo{journal}{Phys. Rev. E} \textbf{\bibinfo{volume}{66}}, \bibinfo{pages}{046132} (\bibinfo{year}{2002}).

\bibitem[{\citenamefont{Lebowitz}(1992)}]{lebowitz}
\bibinfo{author}{\bibfnamefont{S.A.} \bibnamefont{Janowsky}}
\bibnamefont{and} \bibinfo{author}{\bibfnamefont{J.L.} \bibnamefont{Lebowitz}},
\bibinfo{journal}{Phys. Rev. A} \textbf{\bibinfo{volume}{45}}, \bibinfo{pages}{618} (\bibinfo{year}{1992}).

\bibitem[{\citenamefont{Ha}(2002)}]{ha2002PGM}
\bibinfo{author}{\bibfnamefont{M.} \bibnamefont{Ha}}
\bibnamefont{and} \bibinfo{author}{\bibfnamefont{M.} \bibnamefont{den Nijs}},
\bibinfo{journal}{Phys. Rev. E} \textbf{\bibinfo{volume}{66}}, \bibinfo{pages}{036118} (\bibinfo{year}{2002}).

\bibitem[{\citenamefont{Barma}(2006)}]{barma}
\bibinfo{author}{\bibfnamefont{M.} \bibnamefont{Barma}},
\bibinfo{journal}{Physica A}, \textbf{\bibinfo{volume}{372}}, \bibinfo{pages}{22} (\bibinfo{year}{2006}).

\bibitem[{\citenamefont{Krug}(1991)}]{krug}
\bibinfo{author}{\bibfnamefont{J.} \bibnamefont{Krug}},
\bibinfo{journal}{Phys. Rev. Lett.}, \textbf{\bibinfo{volume}{67}}, \bibinfo{pages}{1882} (\bibinfo{year}{1991}).

\bibitem[{\citenamefont{Ha}(2003)}]{ha2003SB}
\bibinfo{author}{\bibfnamefont{M.} \bibnamefont{Ha}},
\bibinfo{author}{\bibfnamefont{J.} \bibnamefont{Timonen}},
\bibnamefont{and} \bibinfo{author}{\bibfnamefont{M.} \bibnamefont{den Nijs}},
\bibinfo{journal}{Phys. Rev. E} \textbf{\bibinfo{volume}{68}}, \bibinfo{pages}{056122} (\bibinfo{year}{2003}).

\bibitem[{\citenamefont{Kolomeisky}(1998)}]{kolomeisky}
\bibinfo{author}{\bibfnamefont{A.~B} \bibnamefont{Kolomeisky}}, \bibinfo{author}{\bibfnamefont{G.~M.} \bibnamefont{Sch{\"u}tz}}, \bibinfo{author}{\bibfnamefont{E.~B} \bibnamefont{Kolomeisky}}, \bibnamefont{and} \bibinfo{author}{\bibfnamefont{J.~P} \bibnamefont{Straley}},
\bibinfo{journal}{J. Phys. A} \textbf{\bibinfo{volume}{31}}, \bibinfo{pages}{6911} (\bibinfo{year}{1998}).

\bibitem[{\citenamefont{Parmeggiani}(2002)}]{parmeggiani}
\bibinfo{author}{\bibfnamefont{A.} \bibnamefont{Parmeggiani}}, \bibinfo{author}{\bibfnamefont{T.} \bibnamefont{Franosch}}, \bibnamefont{and} \bibinfo{author}{\bibfnamefont{E.} \bibnamefont{Frey}},
\bibinfo{journal}{Phys. Rev. E} \textbf{\bibinfo{volume}{70}}, \bibinfo{pages}{046101} (\bibinfo{year}{2004}).

\bibitem[{\citenamefont{Ha}(2002)}]{ha}
\bibinfo{author}{\bibfnamefont{M.} \bibnamefont{Ha}}, \bibinfo{author}{\bibfnamefont{H.} \bibnamefont{Park}}, \bibnamefont{and} \bibinfo{author}{\bibfnamefont{M.} \bibnamefont{den~Nijs}},
\bibinfo{journal}{Phys. Rev. E} \textbf{\bibinfo{volume}{75}}, \bibinfo{pages}{061131} (\bibinfo{year}{2007}).


\bibitem[{\citenamefont{Derrida}(2004)}]{derrida1}
\bibinfo{author}{\bibfnamefont{B.} \bibnamefont{Derrida}}, \bibinfo{author}{\bibfnamefont{B.} \bibnamefont{Doucot}}, \bibnamefont{and} \bibinfo{author}{\bibfnamefont{P.-E} \bibnamefont{Roche}},
\bibinfo{journal}{J. Stat. Phys.} \textbf{\bibinfo{volume}{115}}, \bibinfo{pages}{717} (\bibinfo{year}{2004}); \bibinfo{author}{\bibfnamefont{B.} \bibnamefont{Derrida}}, \bibinfo{author}{\bibfnamefont{J.~L} \bibnamefont{Lebowitz}}, \bibnamefont{and} \bibinfo{author}{\bibfnamefont{E.~R} \bibnamefont{Speer}},
\bibinfo{journal}{J. Stat. Phys.} \textbf{\bibinfo{volume}{107}}, \bibinfo{pages}{599} (\bibinfo{year}{2002}).

\end{thebibliography}
\end{document}